\documentclass[iop,apj,onecolumn]{emulateapj}

\usepackage{graphicx}
\usepackage{psfrag}
\usepackage{amsmath}
\usepackage{bm}
\usepackage{pstricks,etex}

\newcommand{\eq}{Equation~}
\newcommand{\eqs}{Equations~}
\newcommand{\Fig}{Figure~}
\newcommand{\Figs}{Figures~}
\newcommand{\fig}{Figure~}
\newcommand{\figs}{Figures~}

\newcommand{\ie}{i.e.,}
\newcommand{\eg}{{e.g.,}}
\newcommand{\cf}{{cf.}}

\newcommand{\Nu}{\mathrm{Nu}}
\newcommand{\Mu}{\mathrm{nu}}
\newcommand{\Ra}{\mathrm{Ra}}
\renewcommand{\Pr}{\mathrm{Pr}}

\newcommand{\be}{\mathbf{e}}
\newcommand{\bu}{\mathbf{u}}
\newcommand{\nablab}{\bm{\nabla}}
\newcommand{\Tad}{T_{0z}^{\rm ad}}
\newcommand{\Roi}{R_0^{-1}}
\newcommand{\Rci}{R_{\rm c}^{-1}}
\newcommand{\RIi}{R_{\rm I}^{-1}}
\newcommand{\RLi}{R_{\rm L}^{-1}}
\newcommand{\uz}{w}
\newcommand{\ppert}{p}
\newcommand{\Tpert}{T}
\newcommand{\mupert}{\mu}
\newcommand{\rhopert}{\rho}

\newcommand{\tot}{{\rm tot}}
\newcommand{\Tp}{\vartheta}
\newcommand{\Tptot}{\Tp_\tot}
\newcommand{\Ttot}{T_\tot}
\newcommand{\mutot}{\mu_\tot}
\newcommand{\rhotot}{\rho_\tot}
\newcommand{\flux}{{\mathcal F}}

\renewcommand{\aa}{a}
\newcommand{\bb}{b}

\newcommand{\captionsize}{\footnotesize}

\begin{document}

\title{\uppercase{A new model for mixing by double-diffusive convection (semi-convection). II. The transport of heat and composition through layers\vspace{1.5cm}}}
\author{\sc T.~S.~Wood$^1$, P.~Garaud$^1$, and S.~Stellmach$^2$}
\affil{$^1$Department of Applied Mathematics and Statistics, Baskin School of Engineering, University of California Santa Cruz, CA 95064, USA}
\affil{$^2$Institut f\"ur Geophysik, Westf\"alische Wilhelms-Universit\"at M\"unster, M\"unster D-48149, Germany}
\email{tsw25@soe.ucsc.edu}

\begin{abstract}
  Regions of stellar and planetary interiors that are unstable according to the Schwarzschild criterion,
  but stable according to the Ledoux criterion, are subject to a form of oscillatory double-diffusive (ODD)
  convection often called ``semi-convection''.
  In this series of papers, we use an extensive suite of three-dimensional (3D) numerical simulations
  to quantify the transport of heat and composition by ODD convection,
  and ultimately propose a new 1D prescription
  that can be used in stellar and planetary structure and evolution models.
  The first paper in this series demonstrated
  that under certain conditions ODD convection spontaneously transitions
  from an initial homogeneous state of weak wave-breaking turbulence
  into a staircase of fully convective layers,
  which results in a substantial increase in the transport of heat and composition.
  Here, we present simulations of ODD convection in this layered regime,
  we describe the dynamical behavior of the layers,
  and we derive empirical scaling laws for the transport through layered convection.
\end{abstract}


\section{\uppercase{Introduction}}
\label{sec:intro}

A long-standing problem in modeling stellar and planetary interiors
is how to quantify the transport of heat and composition in regions that
have a destabilizing (\ie\ superadiabatic)
thermal stratification and a stabilizing compositional stratification,
such that the total density stratification is stable
according to Ledoux's criterion \citep{SchwarzschildHarm58}.
Such regions are present, for instance,
in giant planets \citep{Stevenson82,LeconteChabrier12},
and in massive stars both main-sequence \citep{Merryfield95}
and post-main-sequence \citep{RobertsonFaulkner72,Langer-etal85,Woosley-etal02}.
It has long been recognized that,
because the molecular diffusivity of composition is typically only a tiny fraction of
the total (radiative plus conductive) diffusivity of temperature,
these regions are subject to a form of double-diffusive instability
\citep{Stern60,Walin64,Veronis65,Kato66}.
The turbulent convection that arises from this instability,
which is often called ``diffusive'' or ``oscillatory'' convection,
enhances the transport of heat and composition,
changing the internal structure of the host object.

Oscillatory double-diffusive convection is closely related to the phenomenon known as ``semi-convection,''
but the latter is often also associated with
the effects of
convective overshoot and the dependence of thermal diffusivity on composition,
neither of which are included in this study.
Instead, we consider a localized region away from any radiative--convective boundary,
in which the thermal and compositional diffusivities can be approximated as constant,
and we study the instability and turbulence arising from the competition between thermal and
compositional stratification.
Following \citet{Paparella-etal02} we refer to convection in this situation as
``oscillatory double-diffusive (ODD) convection.''

The linear instability mechanism that leads to ODD convection can be understood from physical arguments.
In the absence of thermal and compositional diffusion,
vertically displaced parcels of fluid oscillate adiabatically
about their equilibrium position with a frequency determined by the
total density stratification.
But the presence of diffusive processes can amplify the oscillation,
producing an instability that resembles an overstable internal gravity wave \citep{Kato66}.
Indeed, if the thermal diffusivity is sufficiently
large in comparison with the compositional diffusivity, then a displaced parcel
will acquire heat (and therefore entropy) while below its equilibrium position,
and thus will be slightly more
buoyant than its surroundings when it returns to its equilibrium position.
This excess entropy is lost to the cooler fluid above the parcel's equilibrium position,
releasing gravitational potential energy from the superadiabatic stratification.
The amplitude of the oscillation grows exponentially until nonlinear effects are sufficient to
saturate the instability, leading to a state of weakly turbulent ODD convection.

\defcitealias{Mirouh-etal12}{Paper~I}%
After saturation,
ODD convection initially takes the form of a homogenously turbulent state,
in which the time-averaged stratification remains uniform.
Previous work by \citet{Rosenblum-etal11} and \citet{Mirouh-etal12}
has shown that
the turbulent transport of
heat and composition in this homogeneous state is fairly weak.
\citet[][hereafter \citetalias{Mirouh-etal12}]{Mirouh-etal12} derived
empirical formulae for the respective fluxes
from an extensive suite of numerical simulations performed in local Cartesian domains.
However,
in certain cases
the homogeneous state was found to spontaneously transition into
a new state of layered convection, for which the transport of heat and composition is greatly enhanced.
\citetalias{Mirouh-etal12} derived
a simple semi-analytical criterion that predicts under which conditions
spontaneous layering occurs.
Under the physical conditions relevant to planetary and stellar interiors,
spontaneous layering is expected within a significant fraction of the double-diffusive instability range,
specifically in regions that are closer to being Ledoux-unstable (see \fig1 in \citetalias{Mirouh-etal12}).

In this paper, we present further results from numerical simulations
and use them to derive
empirical formulae for the fluxes of heat and composition in layered convection.
We determine how the fluxes depend on the thickness of the layers and on other governing parameters.
We also describe some of the physical properties of layered convection,
paying particular attention to the interfaces between the layers,
and we compare our results with previous studies of ODD convection.
Finally, we discuss our results in the light of recent layered models of giant planets.

\section{The Mathematical Model and Numerical Scheme}
\label{sec:model}

The numerical model that we use to study ODD convection
was described in detail
in \citetalias{Mirouh-etal12}, so here we mention only its most relevant features.
The computational domain is a Cartesian box
with uniform gravity $\mathbf{g} = -g\be_z$.  We solve the evolution equations for
momentum, temperature, and composition in the Boussinesq approximation,
which are
\begin{align}
  \frac{\partial\bu}{\partial t} + \bu\cdot\nablab\bu
    &= -\frac{1}{\rho_0}\nablab\ppert - \frac{\rhopert}{\rho_0}g\be_z + \nu\nabla^2\bu,
    \label{eq:mom-dim} \\
  \frac{\partial \Tpert}{\partial t} + \bu\cdot\nablab \Tpert + (T_{0z} - \Tad)\uz
    &= \kappa_T\nabla^2\Tpert, \\
  \frac{\partial\mupert}{\partial t} + \bu\cdot\nablab\mupert + \mu_{0z}\uz
    &= \kappa_\mu\nabla^2\mupert,
    \label{eq:mu-dim} \\
  \nablab\cdot\bu &= 0.
\end{align}
Here $\ppert$, $\rhopert$, $\Tpert$, and $\mupert$ are perturbations of pressure, density, temperature,
and mean molecular weight respectively,
and $\uz$ is the vertical component of the velocity $\bu$;
all of these quantities are assumed to be periodic in all three Cartesian directions.
We use subscript `0's to denote the unperturbed, background fields, which we assume are functions
of $z$ alone.
The background gradients of temperature, $T_{0z}$, and mean molecular weight, $\mu_{0z}$,
are both negative in the situation considered here, as is the adiabatic temperature gradient,
\begin{equation}
  \Tad = \left(\dfrac{\partial T}{\partial p}\right)_{\!\!\rm ad}p_{0z}.
\end{equation}
We consider a small volume within which the
diffusivities of momentum, $\nu$, temperature, $\kappa_T$, and composition, $\kappa_\mu$,
can be treated as constant, and the density perturbations can be expressed as a linear function of the temperature
and composition perturbations,
\begin{flalign}
  &&\frac{\rhopert}{\rho_0} \;&=\; \beta\mupert - \alpha \Tpert,&
  \label{eq:rho} \\
  &\mbox{where}&
  \alpha = -\frac{1}{\rho_0}\left(\dfrac{\partial\rho}{\partial T}\right)_{\!\!p,\mu}
  \;\;&\mbox{and}\;\;
  \beta = \frac{1}{\rho_0}\left(\dfrac{\partial\rho}{\partial\mu}\right)_{\!\!p,T}.&
\end{flalign}
Both $\alpha$ and $\beta$ are positive for any realistic equation of state.
We approximate $\alpha$, $\beta$, and the other coefficients in \eqs(\ref{eq:mom-dim})--(\ref{eq:mu-dim})
as constant within our small domain,
and we suppose that the background temperature and molecular weight profiles are linear in $z$,
so that the total temperature and composition fields are $\Ttot=\Tpert+T_{0z}z$ and $\mutot=\mupert+\mu_{0z}z$,
with $T_{0z}$, $\Tad$, and $\mu_{0z}$ all constant.
The Schwarzschild and Ledoux stability criteria are then
\begin{flalign}
  &&\alpha(T_{0z} - \Tad) &> 0&
  \label{eq:Schwarz} \\
  &\mbox{and}&\alpha(T_{0z} - \Tad) &> \beta\mu_{0z}&
  \label{eq:Ledoux}
\end{flalign}
respectively.
We are concerned here with regions which satisfy (\ref{eq:Ledoux}) but not (\ref{eq:Schwarz}),
\ie\ regions in which $\Roi > 1$, where $\Roi$ is the ``inverse density ratio,''
\begin{equation}
  \Roi = \dfrac{\beta|\mu_{0z}|}{\alpha|T_{0z}-\Tad|}.
\end{equation}
From a linear stability analysis of \eqs(\ref{eq:mom-dim})--(\ref{eq:mu-dim}),
it can be shown that such regions are unstable to double-diffusive instability if
$\Roi$ lies in the range
\begin{equation}
  1 < \Roi < \Rci
  \label{eq:window}
\end{equation}
\citep{Walin64}
where $\Rci$ is the critical value
\begin{equation}
  \Rci = \dfrac{\kappa_T + \nu}{\kappa_\mu + \nu} = \dfrac{1 + \Pr}{\tau + \Pr}.
\end{equation}
Here, $\Pr$ and $\tau$ are the diffusivity ratios
\begin{flalign}
  && \Pr &= \nu/\kappa_T& \\
  &\mbox{and}& \tau &= \kappa_\mu/\kappa_T.&
\end{flalign}
In astrophysical objects, both $\Pr$ and $\tau$ are typically of order $10^{-2}$ or smaller, and so
the instability window given by \eq(\ref{eq:window}) is rather wide.
The instability is oscillatory, and has a typical horizontal lengthscale
\begin{equation}
  d = \left(\dfrac{\kappa_T\nu}{\alpha g|T_{0z}-\Tad|}\right)^{1/4}
  \label{eq:d}
\end{equation}
\citep[\eg][]{BainesGill69}.

We are interested primarily in the fluxes of heat and composition through ODD convection.
As in \citetalias{Mirouh-etal12} we measure these fluxes in terms of Nusselt numbers, defined as
\begin{flalign}
  &&\Nu_T &= 1 - \dfrac{\left<\uz \Tpert\right>}{\kappa_T(T_{0z}-\Tad)}&
  \label{eq:NuT-dim} \\
  &\mbox{and}&\Nu_\mu &= 1 - \dfrac{\left<\uz\mupert\right>}{\kappa_\mu\mu_{0z}},&
  \label{eq:NuC-dim}
\end{flalign}
where the notation $\big<\cdot\big>$ represents a volume average over the
entire domain.
The total vertical fluxes of heat, $\flux_T$, and composition, $\flux_\mu$, can be reconstructed from these as follows:
\begin{flalign}
  &&\frac{\flux_T}{\rho_0\,c_p} &= \left<\uz \Tpert\right> - \kappa_TT_{0z}& \nonumber \\
  && &= -\Nu_T\kappa_TT_{0z} + (\Nu_T-1)\kappa_T\Tad&
  \label{eq:flux_T} \\
  &\mbox{and}&\flux_\mu &= \left<\uz\mupert\right> - \kappa_\mu\mu_{0z}& \nonumber \\
  && &= -\Nu_\mu\kappa_\mu\mu_{0z},&
  \label{eq:flux_C}
\end{flalign}
where $c_p$ is the specific heat.
We note that, in general, the thermal Nusselt number $\Nu_T$
defined by \eq(\ref{eq:NuT-dim})
is \emph{not} the ratio
of the total heat flux to the diffusive (\ie\ radiative plus conductive) heat flux, unless the adiabatic temperature
gradient $\Tad$ is zero.  However, as in \citetalias{Mirouh-etal12}, we can introduce the
total ``potential temperature''
\begin{align}
  \Tptot &= \Tpert + (T_{0z}-\Tad)z,
\end{align}
which has background gradient $\Tp_{0z} = (T_{0z}-\Tad)$ and perturbation $\Tp=\Tpert$,
in order to write the thermal Nusselt number as
\begin{equation}
  \Nu_T = 1 - \dfrac{\left<\uz\Tp\right>}{\kappa_T\Tp_{0z}}.
\end{equation}
So we may regard $\Nu_T$ as the ratio of the total and diffusive fluxes of potential temperature.
Since $\Tp$ and $\Tpert$ are equal, in what follows we will refer to $\Tp$ as simply ``the temperature perturbation''.

In double-diffusive convection, an important quantity is the ``buoyancy flux ratio'' $\gamma^{-1}$,
which we define as
\begin{align}
  \gamma^{-1} &= \dfrac{\left<g\beta(\uz\mupert - \kappa_\mu\mu_{0z})\right>}
    {\left<g\alpha(\uz\Tp - \kappa_T\Tp_{0z})\right>} \nonumber\\
  &= \dfrac{g\beta\kappa_\mu\mu_{0z}\Nu_\mu}{g\alpha\kappa_T\Tp_{0z}\Nu_T} \nonumber\\
  &= \tau\Roi\dfrac{\Nu_\mu}{\Nu_T} \label{eq:gamma}
\end{align}
\citep{Radko03}\footnote{Unlike \citet{Radko03}, here we define $\gamma^{-1}$ as the ratio of the
\emph{total} fluxes, rather than the turbulent fluxes.  See \citetalias{Mirouh-etal12} for more detail.}.
This is closely related to the core erosion ``efficiency factor'' $\chi$ introduced by \citet{Guillot-etal04},
which measures the fraction of a planet's luminosity that is used to erode the stable compositional
stratification at the boundary with the core.  In the limit where the transport of both heat and composition
is dominated by convection (\ie\ $\Nu_{T,\mu}\gg1$) this fraction is precisely $\gamma^{-1}$.
The gravitational potential energy released from the thermal stratification must exceed the potential
energy used to erode the compositional stratification, which implies that $\gamma^{-1} < 1$.
As shown in \citetalias{Mirouh-etal12},
$\gamma^{-1}$ plays an important role in the formation of layers,
and possibly also controls the evolution of layers after their formation \citep{Radko05}.

For computational convenience, we work mainly with nondimensional quantities.  As in \citetalias{Mirouh-etal12}
we nondimensionalize \eqs(\ref{eq:mom-dim})--(\ref{eq:rho}) using the lengthscale $d$ from \eq(\ref{eq:d})
as well as the scales
\begin{flalign}
  &&[t] &= d^2/\kappa_T,& \\
  &&[\Tp] &= d|T_{0z}-\Tad|,& \\
  &&[\mupert] &= d(\alpha/\beta)|T_{0z}-\Tad|,& \\
  &\mbox{and}&[p] &= \rho_0\dfrac{\nu\kappa_T}{d^2}.&
\end{flalign}
The governing equations are then
\begin{align}
  \frac{1}{\Pr}\left(\frac{\partial\bu}{\partial t} + \bu\cdot\nablab\bu\right)
    &= -\nablab\ppert + (\Tp - \mupert)\be_z + \nabla^2\bu,
    \label{eq:mom} \\
  \nablab\cdot\bu &= 0,
  \label{eq:divu} \\
  \frac{\partial \Tp}{\partial t} + \bu\cdot\nablab \Tp - \uz
    &= \nabla^2\Tp,
    \label{eq:T} \\
  \frac{\partial\mupert}{\partial t} + \bu\cdot\nablab\mupert - \Roi \uz
    &= \tau\nabla^2\mupert,
    \label{eq:mu}
\end{align}
where all variables are now dimensionless.
The definitions (\ref{eq:NuT-dim})--(\ref{eq:NuC-dim})
of the Nusselt numbers then become
\begin{flalign}
  &&\Nu_T &= 1 + \left<\uz \Tp\right>&
  \label{eq:NuT} \\
  &\mbox{and}&\Nu_\mu &= 1 + \dfrac{\left<\uz\mupert\right>}{\tau\Roi}.&
  \label{eq:NuC}
\end{flalign}
We will also make use of alternative definitions of the Nusselt numbers,
\begin{flalign}
  &&\Mu_T &= 1 + \left<|\nablab \Tp|^2\right>&
  \label{eq:MuT} \\
  &\mbox{and}&\Mu_\mu &= 1 + \dfrac{\left<|\nablab\mupert|^2\right>}{(\Roi)^2},&
  \label{eq:MuC}
\end{flalign}
which are based on the rates of thermal and compositional dissipation,
and are sometimes called ``Cox numbers''.
We note that $\Mu_T$ and $\Mu_\mu$ are guaranteed to be positive at all times,
whereas $\Nu_T$ and $\Nu_\mu$ are not.
However, it can be readily verified, from \eqs(\ref{eq:divu})--(\ref{eq:mu}),
that the definitions (\ref{eq:NuT})--(\ref{eq:NuC}) and (\ref{eq:MuT})--(\ref{eq:MuC})
yield the same result after taking a time average in a statistically steady state
\citep{Malkus54,ShraimanSiggia90}.
The advantage of retaining both forms is explained in Section~\ref{sec:measure}.

\section{TYPICAL PROPERTIES OF LAYERED CONVECTION}
\label{sec:typical}

In principle,
any particular instance of ODD convection is characterized by the three dimensionless parameters
$\Roi$, $\Pr$, and $\tau$.
However,
in \citetalias{Mirouh-etal12} it was shown that, within the instability window given by \eq(\ref{eq:window}),
there is another critical value of $\Roi$, say $\RLi$, below which ODD convection spontaneously transitions into
layered convection.
\citetalias{Mirouh-etal12} presented empirical scaling laws for $\RLi$, and for the transport of heat and composition
prior to the development of layers.  It was shown that, in the parameter regime relevant to astrophysical objects,
for which $\Pr \simeq \tau \ll 1$, the critical value $\RLi \sim \Pr^{-1/2}$.
This result implies that there is a significant range of density ratios $\Roi$ for which
layers form spontaneously, and so the transport through layered convection must be considered
in models of astrophysical objects.\footnote{This is in contrast to the ``fingering'' regime
of double-diffusive convection, for which spontaneous layer formation is not expected
at astrophysical parameter values \citep{Traxler-etal11a},
unless a different mechanism of layer formation operates
\citep[][to appear in {\sl The Astrophysical Journal}]{Brown-etal13}.}
A preliminary study of layered convection by \citet{Rosenblum-etal11} demonstrated that
the turbulent transport is strongly dependent on the layer height, $H$ say, in addition to $\Pr$, $\tau$,
and $\Roi$.  Their results suggest that $\Nu_T$ and $\Nu_\mu$ both follow a power law in $H$,
with an exponent between $1$ and $4/3$.
In what follows we present a more comprehensive study of the dependence of the turbulent fluxes on
each of the parameters $H$, $\Pr$, $\tau$, and $\Roi$.

\begin{figure}[h]
  \centering%
  \includegraphics[width=16cm]{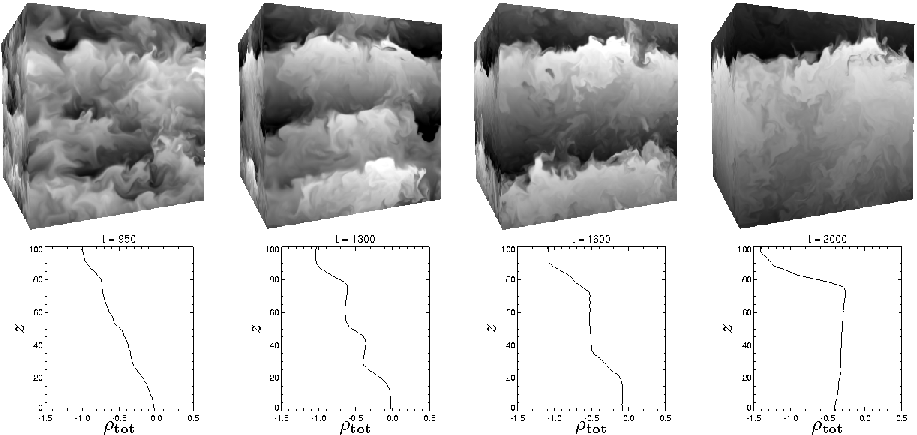}%
  \caption{Composition perturbation $\mupert$ (top row)
    and potential density profiles $\rhotot(z)$ (bottom row)
    at four different times in a simulation with $\Roi = 1.5$, $\Pr=\tau=0.03$, and a domain size of $(100d)^3$.
    The first column shows the homogeneous phase of ODD convection prior to layer formation.
    The remaining columns show layered convection with 3, 2, and 1 layers respectively.
    Because the perturbation is plotted in the top row, rather than the total composition,
    the layers appear as regions with a
    relatively weak vertical gradient, and the interfaces appear as regions with a strong vertical gradient
    of the opposite sign.
    The color bar for each plot is scaled to the maximum of the perturbation $\mupert$,
    which increases as the layers merge.
    In the profiles, $\rhotot$ has been rescaled such that the total change in $\rhotot$
    across the domain is unity.}
  \label{fig:evolution}
\end{figure}

In order to provide a clear description of the transport properties of layered convection,
and the subtleties involved in measuring the thermal and compositional fluxes in the layered phase,
we first revisit one particular simulation that was previously
presented in \citetalias{Mirouh-etal12}.
This simulation has $\Roi = 1.5$, $\Pr=0.03$, $\tau=0.03$, and a
domain size of $(100d)^3$.
As described in \citetalias{Mirouh-etal12}, this simulation spontaneously transitions from
homogeneous ODD convection to layered convection, and the three layers that initially form subsequently
merge into two layers, then into one layer.  This evolution is illustrated in \fig\ref{fig:evolution},
which shows snapshots of the composition perturbation $\mupert$ during each phase of the simulation,
as well as profiles of the horizontally-averaged total ``potential density,''
\begin{equation}
  \rhotot = \rho_0(\beta\mutot - \alpha\Tptot).
  \label{eq:rhopot}
\end{equation}
The profiles show that
$\rhotot$ is roughly constant
within each layer, but increases slightly with height.
The local Ledoux stability criterion can be expressed as $\dfrac{\partial\rhotot}{\partial z} < 0$
(\cf\ \eq(\ref{eq:Ledoux})),
so this indicates that the layers are
fully convective.
Within each layer, the total composition $\mutot$ and the total potential temperature $\Tptot$
are both roughly uniform, and there are strong jumps in $\mutot$ and $\Tptot$ across the interfaces
between adjacent layers.

Spontaneous layer formation
similar to that observed here
has previously been reported in simulations of double-diffusive
convection in the ``fingering'' regime, for which the thermal stratification is stabilizing and the compositional
stratification is destabilizing \citep[\eg][]{Stellmach-etal11},
but only for Prandtl numbers $\Pr > 1$ \citep{Traxler-etal11a}.
However, in fingering convection the initial height of the layers
is typically much larger than the lengthscale of the linear
double-diffusive instability, whereas here we find that layers initially form
on a scale comparable to that of the linear instability.
In the simulation shown in \fig\ref{fig:evolution}, for example,
the fastest growing linear mode has a wavelength of roughly $20d$,
\ie\ $1/5$ of the domain size, and the initial layer height is $1/3$ of the domain size.

\begin{figure}[h]
  \centering%
  \includegraphics[width=16cm]{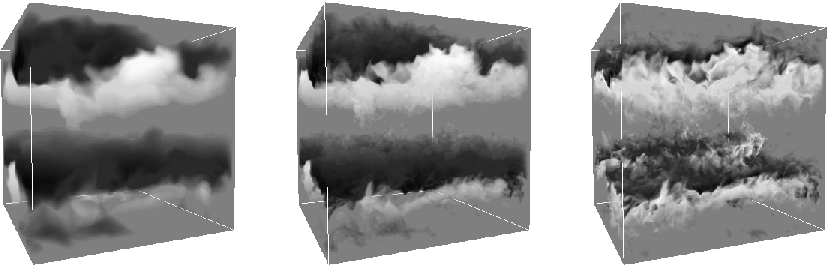}%
  \caption{Perturbations of temperature (left), composition (center), and density (right) from the simulation shown in
  \fig\ref{fig:evolution}, at $t = 1600$.
  Regions where the perturbations are small have been made transparent in order to show
  the interfaces in more detail.}
  \label{fig:interface}
\end{figure}

\Fig\ref{fig:interface} shows more detail of the interface structure during the 2-layer phase.
We find that the interfaces are highly dynamic and turbulent, exhibiting structure on scales
that are much smaller than the wavelength of the fastest growing linear mode.
This is in marked contrast to the flat, laminar interfaces found in simulations and laboratory experiments
of ODD convection at larger Prandtl number
\citep[\eg][both of which have $\Pr\simeq7$]{NoguchiNiino10a,Carpenter-etal12b}.
The composition field exhibits structure on finer scales than the
temperature field, as expected because of the smaller diffusivity of composition.
On larger scales, the temperature and composition fields display similar features,
such that the temperature field resembles a coarse-grained version of the composition field.
By contrast, the density field has even finer structure, since it depends on the difference between the
composition and temperature fields.

\begin{figure}[h]
  \centering%
  \includegraphics[width=10cm]{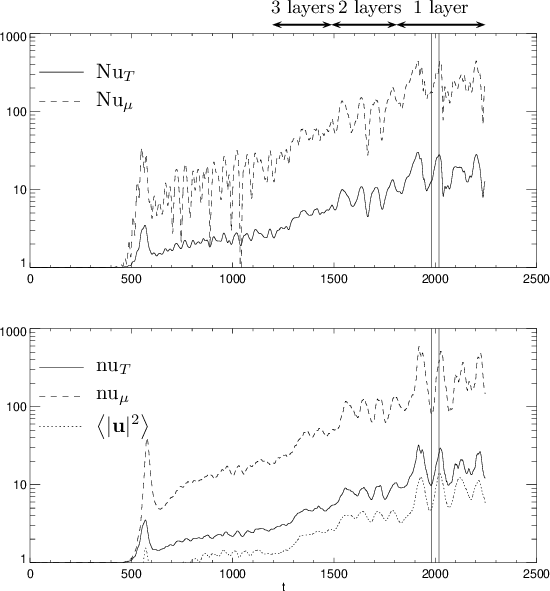}%
  \caption{The Nusselt numbers, defined by \eqs(\ref{eq:NuT})--(\ref{eq:MuC}),
  and the mean kinetic energy in the simulation
  shown in \figs\ref{fig:evolution} and \ref{fig:interface}.
  The initial linear instability grows exponentially until $t\approx570$, then saturates, leading
  to homogenous ODD convection.
  Layers first form at $t\approx1200$; the duration of the 3 layer, 2 layer, and 1 layer phases
  are indicated at the top of the plot.
  The vertical lines indicate the two times shown in \fig\ref{fig:min_max}.
  Using $\Mu_{T,\mu}$ rather than $\Nu_{T,\mu}$ to measure the mean fluxes
  has the advantage of filtering out fluctuations
  in the fluxes on short timescales, while retaining the important long timescale information.}
  \label{fig:Nu}
\end{figure}

The fluxes
through this simulation, measured in terms of Nusselt numbers $\Nu_{T,\mu}$ and $\Mu_{T,\mu}$,
are plotted in \fig\ref{fig:Nu} as a function of time, along with the mean-squared velocity,
$\left<|\mathbf{u}|^2\right>$.
As mentioned earlier, the mean value of each flux increases significantly when layers first form,
and increases still further when layers merge \citep{Rosenblum-etal11}.
The mean kinetic energy also increases, roughly in proportion
with the fluxes, indicating that the simulation becomes more turbulent as the layers become thicker.
We note that the two definitions for the Nusselt numbers both yield consistent results for the mean flux,
although the definition based on the instantaneous fluxes exhibits larger fluctuations in the
homogeneous, pre-layer
phase of the simulation, during which the dynamics are dominated by wave-like,
oscillatory
motions \citep{Mirouh-etal12}.

\begin{figure}[h]
  \centering%
  \includegraphics[height=5cm]{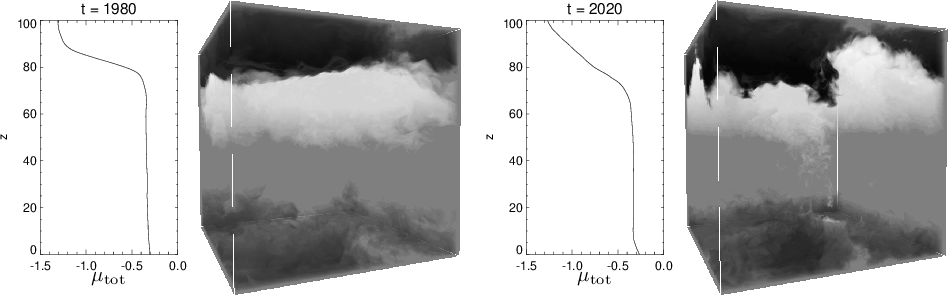}%
  \caption{The interface structure at times $t=1980$ and $t=2020$ during the one-layer phase.
  The profiles show the horizontal average of the total composition $\mutot$, and the volume rendering shows
  the perturbation $\mupert$ using the same color bar at both times.}
  \label{fig:min_max}
\end{figure}

Between layer merger events
the mean fluxes and kinetic energy oscillate
quasi-periodically between high and low values, with relative departures from the mean flux of order unity.
The oscillations occur as the flow in the simulation cycles between states that are highly turbulent and states
that are more quiescent.
In the quiescent state, the interfaces are roughly horizontal, whereas in the turbulent state the interfaces
have undulations on a horizontal scale comparable to the thickness of the layers.
For illustration, \fig\ref{fig:min_max} shows
the composition perturbation, and the horizontally-averaged total composition $\mutot$,
at two times
during the one-layer phase.
At the first time, $t = 1980$, the interface is in a quiescent state, and at the second,
$t = 2020$, the interface is in a more turbulent state.
These two times correspond to minima and maxima in the Nusselt numbers,
as shown in \fig\ref{fig:Nu}.

The quasi-periodic nature of the oscillations in the fluxes suggests that they
are caused by large-scale gravity-wave
motions of the interfaces.  For an interfacial gravity wave of horizontal scale $H$, the oscillation frequency is
\begin{equation}
  \omega \simeq \sqrt{\dfrac{\Delta\rhotot}{\rho_0}\dfrac{g}{H}}
\end{equation}
\citep{Rayleigh83}, where $\Delta\rhotot$ represents the jump in potential density across the interface.
For a staircase of layers of thickness $H$ we have $\Delta\rhotot = H\rho_0|\beta\mu_{0z} - \alpha\Tp_{0z}|$,
and so the characteristic gravity-wave frequency is
\begin{align}
  \omega \simeq \sqrt{g|\beta\mu_{0z} - \alpha\Tp_{0z}|}
  = N,
\end{align}
where $N$ is the buoyancy frequency associated with the mean stratification,
which in dimensionless units is $N = \sqrt{\Pr(\Roi-1)}$.
In the simulation shown in \fig\ref{fig:Nu} this corresponds to a (dimensionless) oscillation period of
$2\pi/\omega \simeq 50$, which is approximately half of the typical period of the observed flux oscillations.
We emphasize, however, that the disturbances to the interfaces seen in \fig\ref{fig:min_max}
are far more complex in structure than linear gravity waves.
Moreover, the large-scale motions of the interfaces are not strictly periodic in time --- they grow and decay
cyclically, but they do not oscillate about zero.

Previous models of layered convection in astrophysical objects \citep[\eg][]{Spruit92}
have assumed that the interfaces are horizontal, laminar, and diffusive.
If that were the case, the fluxes though the interfaces could be determined
simply by measuring the vertical gradients of
the horizontally-averaged $\Ttot$ and $\mutot$ profiles
in the center of each interface.
\Figs\ref{fig:Nu} and \ref{fig:min_max} show that this type of model cannot be valid here,
because the fluxes are maximal when these gradients are weakest.
In fact, since the interfaces are not horizontal, the fluxes depend on the local gradients within
the interfaces \emph{and} on the effective surface area of the interfaces.
\Fig\ref{fig:min_max} suggests that variations in the surface area
are the main cause of the oscillations in the fluxes.

The interface behavior seen here somewhat resembles
that seen in the compressible, two-dimensional simulations
of \citet{Biello01} at low Prandtl number.
\citeauthor{Biello01} also attributed the transport across the interfaces to the breaking of interfacial gravity waves,
although the interpretation was complicated by interactions with the boundaries of the domain.
The disturbances to the interface in \fig\ref{fig:min_max} have the same horizontal lengthscale
as the width of the domain, $100d$, suggesting that, here also, the dynamics may be sensitive to the choice
of domain size and the periodic boundary conditions.
However, as described in Appendix~\ref{sec:robust},
similar behavior is seen in all of our layered simulations,
and the results do not depend significantly on the size of the domain.

\section{MEAN FLUXES THROUGH LAYERED CONVECTION}

\subsection{Method for Measuring the Mean Fluxes}
\label{sec:measure}

Our aim is to determine how the mean fluxes of heat and composition,
measured in terms of dimensionless Nusselt numbers, depend on the dimensionless
parameters $R_0^{-1}$, $\Pr$, $\tau$,
and the average (dimensionless) layer height, which we denote as $H$.
In making this determination, we are implicitly assuming that each layered phase
is a statistically steady state, so that the mean fluxes in each phase are well defined.
This point of view is based on the observation of step-like increases in the fluxes seen,
for example, in \fig\ref{fig:Nu},
although strictly speaking the occurrence of layer mergers implies that each phase is
not truly statistically steady.
Moreover, even in the absence of layer mergers, the possibility of long-timescale transients
in the turbulent fluxes cannot be ruled out \citep[\eg][]{Ahlers-etal06,BrownAhlers06}.
One way to check that the mean fluxes we measure are meaningful and reproducible properties
of our model is to compare the fluxes in simulations with the same physical parameters
but different numerical resolutions, computational domain sizes, and initial conditions.
We have performed many such tests, several of which are described in Appendix~\ref{sec:robust},
to ensure that our flux measurements are robust.

As described in the last section, during each layered phase
the fluxes of heat and composition oscillate quasi-periodically about a mean value,
and reliably calculating the mean fluxes requires time averaging over several of these
oscillations.
Unfortunately, the duration of each layered phase is typically not much longer
than the period of the oscillations (see \fig\ref{fig:Nu}, for example), and so even after time averaging,
our measurements of the mean flux still have significant errors.
Only in the one-layer phase, for which merging is artificially suppressed by the periodic boundary
conditions, can averaging be performed over many oscillations.
In order to make the best use of all available data,
when measuring the mean fluxes we also
estimate the error in our measurements.
These errors are used to weight the measurements when fitting to empirical scaling laws in Section~\ref{sec:results}.
The mean fluxes
and errors are calculated
as follows:
\begin{enumerate}
  \item We first identify the number of layers during each phase of the simulation, by counting the number of ``steps''
    in the vertical profiles of the horizontally averaged fields,
    and matching each merger event with the corresponding increase in the Nusselt numbers.
  \item Once we have identified the time interval corresponding to a particular layered phase, we then
    divide this interval into four equal subintervals, and calculate
    the average values of $\Nu_T$, $\Nu_\mu$, $\Mu_T$, and $\Mu_\mu$
    during each subinterval using \eqs(\ref{eq:NuT})--(\ref{eq:MuC}).
    Thus we obtain eight separate measurements of the thermal and compositional fluxes during each layered phase.
  \item We treat each of our eight measurements as independent, and calculate their mean and standard deviation,
    which yields estimates for the mean flux and the measurement error.
    (A similar method was used to estimate the measurement error in \citetalias{Mirouh-etal12}.)
  \item In the same fashion, we calculate the buoyancy flux ratio $\gamma^{-1}$
    for each quarter interval using \eq(\ref{eq:gamma}), and for both definitions of the Nusselt numbers.
    As before, we use the resulting eight measurements to estimate the mean value of $\gamma^{-1}$
    and our measurement error.
\end{enumerate}

We have applied this procedure to each of the simulations from \citetalias{Mirouh-etal12}
in which layers formed spontaneously,
as well as to many additional simulations that have since been performed.
The parameters of each simulation,
including those from \citetalias{Mirouh-etal12} that developed layers,
are listed in Table~\ref{tab:runs}.
\begin{deluxetable*}{cccccc}

\tablecaption{Parameters of each simulation}

\tablehead{\colhead{$\Pr$} & \colhead{$\tau$} & \colhead{$\Roi$} & \colhead{Domain size} & \colhead{\begin{tabular}{c}Resolution\\ (Fourier modes)\end{tabular}} & \colhead{Initial layer No.}}

\startdata
\sl 0.1 & \sl 0.1 & \sl 1.1 & \sl 100$^3$ & \sl 128$^3$ & \sl 2 \\
0.1 & 0.1 & 1.1 & 100$^3$ & 64$^3$ & 2 \\
0.1 & 0.1 & 1.1 & 50$^3$ & 64$^3$ & 1 \\
\sl 0.3 & \sl 0.3 & \sl 1.1 & \sl 100$^3$ & \sl 128$^3$ & \sl 2 \\
0.3 & 0.3 & 1.1 & 100$^3$ & 64$^3$ & 2 \\
\sl 0.3 & \sl 0.3 & \sl 1.15 & \sl 100$^3$ & \sl 64$^2$$\times$128 & \sl 2 \\
0.3 & 0.3 & 1.15 & 100$^3$ & 64$^3$ & 2 \\
0.3 & 0.3 & 1.15 & 200$^2$$\times$100 & 128$^3$ & 2 \\
0.3 & 0.3 & 1.15 & 100$^2$$\times$200 & 64$^2$$\times$256 & 4 \\
0.33 & 0.33 & 1.15 & 200$^2$$\times$400 & 128$^2$$\times$512 & 8 \\
0.1 & 0.1 & 1.2 & 100$^3$ & 64$^3$ & 2 \\
0.1 & 0.1 & 1.2 & 200$^2$$\times$100 & 128$^2$$\times$64 & 2 \\
0.3 & 0.3 & 1.2 & 100$^3$ & 32$^3$ & 3 \\
\sl 0.3 & \sl 0.3 & \sl 1.2 & \sl 100$^3$ & \sl 64$^3$ & \sl 2 \\
0.3 & 0.3 & 1.2 & 178$^3$ & 128$^3$ & 3 \\
0.3 & 0.3 & 1.2 & 89$^2$$\times$178 & 64$^2$$\times$128 & 4 \\
0.3 & 0.3 & 1.2 & 89$^2$$\times$178 & 32$^2$$\times$64 & 4 \\
0.3 & 0.3 & 1.2 & 200$^2$$\times$100 & 128$^3$ & 2 \\
\sl 0.1 & \sl 0.1 & \sl 1.25 & \sl 100$^3$ & \sl 128$^3$ & \sl 2 \\
0.1 & 0.1 & 1.25 & 100$^3$ & 64$^3$ & 2 \\
0.1 & 0.1 & 1.25 & 50$^2$$\times$100 & 64$^2$$\times$128 & 3 \\
\sl 0.3 & \sl 0.3 & \sl 1.25 & \sl 100$^3$ & \sl 64$^3$ & \sl 2 \\
\sl 0.01 & \sl 0.01 & \sl 1.5 & \sl 100$^3$ & \sl 192$^3$ & \sl 2 \\
\sl 0.03 & \sl 0.03 & \sl 1.5 & \sl 100$^3$ & \sl 192$^2$$\times$256 & \sl 3 \\
0.03 & 0.03 & 1.5 & 50$^3$ & 96$^3$ & 2 \\
0.03 & 0.03 & 1.5 & 50$^3$ & 64$^3$ & 2 \\
0.03 & 0.03 & 1.5 & 50$^3$ & 48$^3$ & 2 \\
0.03 & 0.03 & 1.5 & 50$^3$ & 128$^3$ & 2 \\
\sl 0.1 & \sl 0.1 & \sl 1.5 & \sl 100$^3$ & \sl 64$^3$ & \sl 2 \\
\sl 0.01 & \sl 0.01 & \sl 2.0 & \sl 100$^3$ & \sl 192$^3$ & \sl 3 \\
\sl 0.3 & \sl 0.03 & \sl 1.1 & \sl 100$^3$ & \sl 192$^3$ & \sl 3 \\
\sl 0.3 & \sl 0.1 & \sl 1.1 & \sl 100$^3$ & \sl 128$^3$ & \sl 2 \\
0.3 & \sl 0.1 & \sl 1.1 & \sl 100$^3$ & \sl 64$^3$ & \sl 2 \\
\sl 0.03 & \sl 0.3 & \sl 1.1 & \sl 100$^3$ & \sl 192$^3$ & \sl 1 \\
\sl 0.1 & \sl 0.3 & \sl 1.1 & \sl 100$^3$ & \sl 64$^3$ & \sl 2 \\
\sl 0.3 & \sl 0.1 & \sl 1.2 & \sl 100$^3$ & \sl 80$^3$ & \sl 2 \\
\sl 0.03 & \sl 0.3 & \sl 1.2 & \sl 100$^3$ & \sl 128$^3$ & \sl 2 \\
\sl 0.1 & \sl 0.3 & \sl 1.2 & \sl 100$^3$ & \sl 80$^3$ & \sl 2 \\
\sl 0.3 & \sl 0.03 & \sl 1.25 & \sl 100$^3$ & \sl 128$^3$ & \sl 4 \\
\sl 0.3 & \sl 0.1 & \sl 1.4 & \sl 100$^3$ & \sl 64$^3$ & \sl 3 \\
\sl 0.3 & \sl 0.03 & \sl 1.5 & \sl 100$^3$ & \sl 128$^3$ & \sl 3
\enddata

\tablecomments{In all runs, layers eventually merge into one layer.
Cases already presented in \citetalias{Mirouh-etal12} are printed in italics.
The numerical resolution is given in terms of the number of Fourier modes;
the number of grid points in each dimension is larger by a factor of three.}

\label{tab:runs}
\end{deluxetable*}

We present below
the results of these measurements
and describe their dependence on
the parameters $\Pr$, $\tau$, $\Roi$, and $H$.
For reasons explained in the Appendix, only the flux measurements
from simulations performed in domains of width $\geqslant100d$
are presented below.

We note that the spontaneous emergence of layers described in \citetalias{Mirouh-etal12}
is not the only mechanism by which layers can develop in ODD convection.
Laboratory experiments
and observations of Earth's oceans indicate that
the formation of layers can be ``triggered''
by external influences
\citep{TurnerStommel64,Kelley-etal03}.
Layered convection might even be metastable under conditions
for which the linear ODD instability is absent
\citep{Veronis65,Spiegel72}.
In this paper, however, we only consider the regime in which layers form spontaneously, $1 < \Roi < \RLi$,
in order to guarantee the reproducibility of our results.

\subsection{Parameter Dependence of the Mean Fluxes}
\label{sec:results}

\begin{figure}[h!]
  \centering%
  \includegraphics[width=17cm]{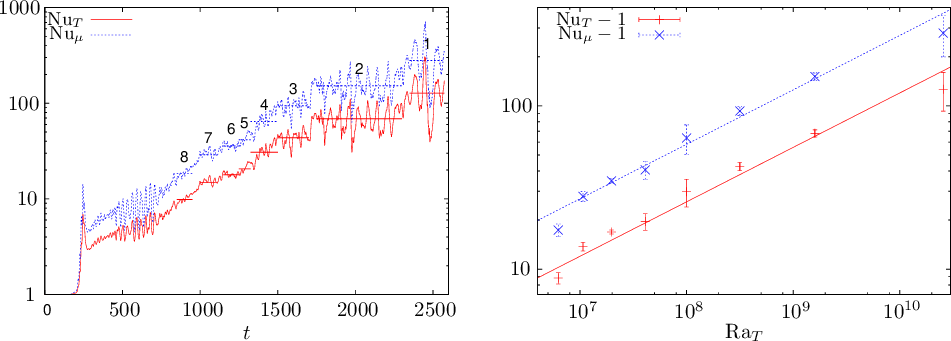}%
  \caption{\captionsize
    a) Time series of the Nusselt numbers from a simulation with $\Pr=\tau=0.33$, $\Roi=1.15$, and
    domain size $200^2\times400$.
    The number of layers in each layered phase is indicated on the plot.
    The horizontal lines indicate the approximate duration of each layered phase and the mean
    flux determined by the method described in Section~\ref{sec:measure}.
    b) Dependence of $\Nu_T$ and $\Nu_\mu$ on $\Ra_T$ in the same simulation.
    The errors in the measurements were determined by the method described in Section~\ref{sec:measure}.
    The solid and dashed lines both have a slope of $1/3$.
  }
  \label{fig:Nu_tall_box}
\end{figure}

To facilitate comparison of our results with those of previous studies, we introduce
the thermal Rayleigh number for layered convection, which is
\begin{align}
  \Ra_T(H) = \dfrac{g\alpha |T_{0z}-\Tad|(Hd)^4}{\kappa_T\nu}
  = H^4,
  \label{eq:RaT}
\end{align}
where $Hd$ is the (dimensional) layer height.
Owing to computational limitations, our results cover only a limited range
of $\Pr$ and $\tau$ values.  Moreover, by choosing to analyze only those simulations
in which layers form spontaneously, we are restricted to a rather narrow range for $\Roi$
\citepalias[see][]{Mirouh-etal12}.
On the other hand, the range of $\Ra_T$ values that can be achieved in each simulation is limited only
by the height of the computational domain.  Moreover, the fact that layers merge allows
us to test the dependence of the Nusselt numbers on $\Ra_T$
using data from a single simulation.
The tallest simulation presented here has a domain of height $400d$, and initially formed eight layers.
\Fig\ref{fig:Nu_tall_box} shows the Nusselt numbers in this simulation, plotted as functions of
time and $\Ra_T$.
We find that both $\Nu_T$ and $\Nu_\mu$ are well approximated by a power law in  $\Ra_T$
with an exponent of $1/3$.
This result can be explained physically
if we suppose that the mean flux depends only on the jumps in potential temperature and composition
across the interfaces, $\Delta\Tp$ and $\Delta\mu$, say, and not on any properties of the layers.
More precisely, let us suppose that the (dimensional) buoyancy fluxes $g\alpha\kappa_T\Tp_{0z}\Nu_T$ and
$g\beta\kappa_\mu\mu_{0z}\Nu_\mu$ depend only on the buoyancy jumps $g\alpha\Delta\Tp$ and
$g\beta\Delta\mu$ and the diffusivities $\kappa_T$, $\kappa_\mu$, and $\nu$.
On dimensional grounds, the fluxes must then be proportional to $(\Delta\Tp)^{4/3}$ \citep{Turner65}.
In our simulations, the \emph{total} change in temperature between the top and bottom of the domain
is fixed, and so the jump across each interface is inversely proportional to the number of layers, and
therefore proportional to the height of the layers.  The exponent $1/3$ then follows from equation~(\ref{eq:RaT}).

We now ask whether \emph{all} of our Nusselt number measurements can be fitted to the same
power law.  In order to fit all of the data simultaneously, we must make some assumption regarding the
dependence not only on $\Ra_T$, but also on $\Pr$, $\tau$, and $\Roi$.
We consider the thermal and compositional Nusselt numbers in turn,
and then the flux ratio $\gamma^{-1}$.

\subsubsection{The Heat Flux ($\Nu_T$)}

\begin{figure}[h!]
  \centering%
  \includegraphics[width=\textwidth]{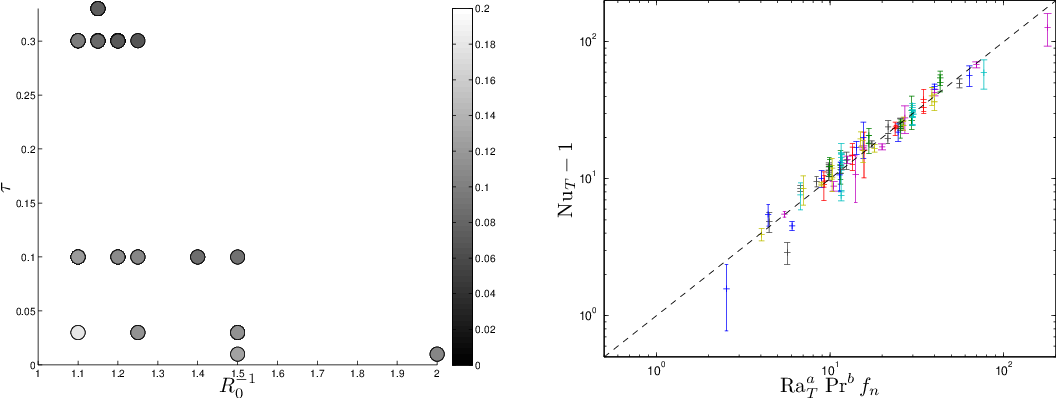}%
  \caption{\captionsize
    Left panel:
    The best-fit values for the coefficients $f_n$ defined in \eq(\ref{eq:NuT_guess2}).
    Right panel:
    Comparison between the measured values of $\Nu_T - 1$ and the best fit for the formula
    in \eq(\ref{eq:NuT_guess2}) with exponents $a=0.34$ and $b=0.34$.
    In the online version of this figure, data from different parameter pairs $(\Roi,\tau)$
    are plotted in different colors.}
  \label{fig:NuT_fit}
\end{figure}

Many analytical and heuristic arguments have been made that the turbulent transport of heat through
layered convection follows a power law in both the Rayleigh number, $\Ra_T$, and the
Prandtl number, $\Pr$ \citep[e.g.,][]{Spruit92,Balmforth-etal06}.
But its dependence on
the parameters $\Roi$ and $\tau$ is less clear.
As a first step toward fitting our data
to an analytical formula for $\Nu_T$, we therefore assume a dependence of the form
\begin{equation}
  \Nu_T - 1 \;\; = \;\; \Ra_T^\aa\;\Pr^\bb\,f(\Roi,\tau),
  \label{eq:NuT_guess}
\end{equation}
where the exponents $\aa$ and $\bb$, and the function $f$, are to be determined.
We group all of our $\Nu_T$ measurements according to their values of $\Roi$ and $\tau$,
which leads to 15 distinct groups:
$(\Roi,\tau) = (1.1,0.03)$, $(1.1,0.1)$, $(1.1,0.3)$, $(1.15,0.3)$, $(1.15,0.33)$, $(1.2,0.1)$, $(1.2,0.3),
(1.25,0.03)$, $(1.25,0.1)$, $(1.25,0.3)$, $(1.4,0.1)$, $(1.5,0.01)$, $(1.5,0.03)$, $(1.5,0.1)$, $(2, 0.01)$.
We then look for a fit of the form
\begin{equation}
  \Nu_T - 1 \;\; = \;\; \Ra_T^\aa\;\Pr^\bb\,f_n,
  \label{eq:NuT_guess2}
\end{equation}
where $n = 1, ..., 15$ represents the index of each group.
To determine the coefficients
$\aa$, $\bb$, and $f_n$
we perform a weighted least-squares minimization
with weights proportional to the inverse square of the estimated error
in the measurements of $\Nu_T$.
The best fit to the data has exponents $\aa = 0.34\pm0.01$ and $\bb = 0.34\pm0.03$.
The best fit values for the coefficients $f_n$ are shown in \fig\ref{fig:NuT_fit},
along with a comparison between the data and the fit.
Most of the data points lie within one standard error of the best fit line, although there are significant
outliers for both low and high values of $\Nu_T$.
The lowest values of $\Nu_T$ are found in simulations that are only weakly turbulent,
for which we would not expect a simple scaling law to hold.
The presence outliers at the highest values of $\Nu_T$, on the other hand,
may result from insufficient time-averaging or numerical resolution
in the simulations with the tallest layers.

The fact that the best fit has exponents $\aa\approx\bb$ suggests that the heat flux depends only
on the parameters $\Roi$, $\tau$, and the product $\Ra_T\Pr$.
In that case the heat flux is independent of viscosity,
as predicted by \citet{Spruit92} for layered convection
with $\Ra_T\gg1$ and $\Pr < 1$.
However, the model of \citet{Spruit92} predicts exponents $\aa=\bb=1/4$, rather than $1/3$,
which is incompatible with our results.

\begin{figure}[h!]
  \centering%
  \includegraphics[height=8cm]{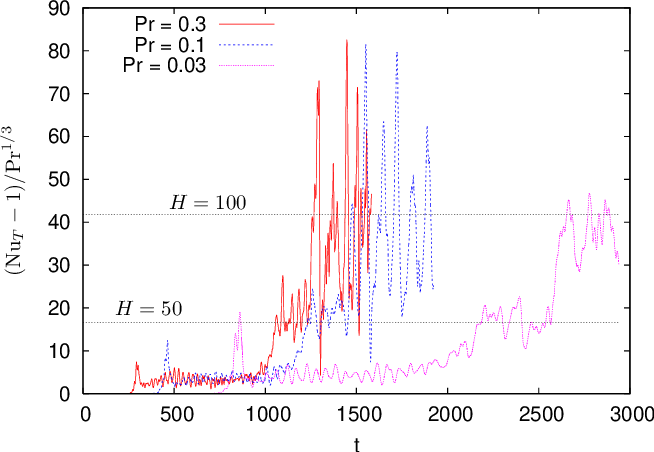}%
  \caption{\captionsize
    Time series of
    the turbulent heat flux,
    rescaled by $\Pr^{-1/3}$,
    in three simulations
    with $\tau = 0.3$, $R_0^{-1}=1.2$, and
    domain height $100d$, but different values of $\Pr$.
    The dashed lines indicate the flux predicted by the best fit
    for the two-layer and one-layer phases, $0.09H^{4/3}$.}
  \label{fig:depend_nu}
\end{figure}

Using the result that $\Nu_T-1 \propto \Pr^{1/3}$ we can directly compare
simulations performed with different values of $\Pr$ but the same values of $\tau$ and $R_0^{-1}$.
An example is given in \fig\ref{fig:depend_nu},
which overlays time series of $\Nu_T$ in three simulations that each have $\tau = 0.3$, $R_0^{-1}=1.2$,
and domain size $(100d)^3$.  For these parameters, the best fit prediction is $\Nu_T - 1 = 0.09H^{4/3}\Pr^{1/3}$.
Since each of these simulations has the same (dimensionless) domain height, we can directly compare their
two-layer ($H=50$) and one-layer ($H=100$) phases;
we find that, as predicted, $(\Nu_T-1)/\Pr^{1/3}$ has the same dependence on $H$ in each of these simulations.

The best-fit values for the coefficients $f_n$, presented in \fig\ref{fig:NuT_fit},
show a weak but well-defined dependence on $\Roi$ and $\tau$
(for fixed $\Ra_T$ and $\Pr$).
In particular, we find that
decreasing either the composition gradient $\mu_{0z}$ or the compositional diffusivity $\kappa_\mu$
produces a slight increase in the mean heat flux.
We interpret this result as evidence that changing $\mu_{0z}$ and $\kappa_\mu$
affects the heat flux only indirectly, by changing the overall level of turbulence.

\subsubsection{The Composition Flux ($\Nu_\mu$)}
We apply the same analysis for the compositional flux, \ie\ we suppose that
\begin{equation}
  \Nu_\mu - 1 \;\; = \;\; \Ra_T^\aa\;\Pr^\bb\,f_n,
  \label{eq:NuC_guess2}
\end{equation}
and fit the data to determine $\aa$, $\bb$, and $f_n$.
This time the best-fit values for the exponents are $\aa = 0.37\pm0.01$ and $\bb = 0.27\pm0.04$.
This is roughly consistent with the trend observed in \fig\ref{fig:Nu_tall_box}b,
but indicates that the compositional flux, unlike the heat flux, depends on viscosity
even for $\Pr\ll1$.
This is perhaps not surprising given that the composition field is more sensitive
than the temperature field
to the small-scale motions of the fluid (\eg\ see \fig\ref{fig:interface}).

\begin{figure}[h!]
  \centering%
  \includegraphics[width=\textwidth]{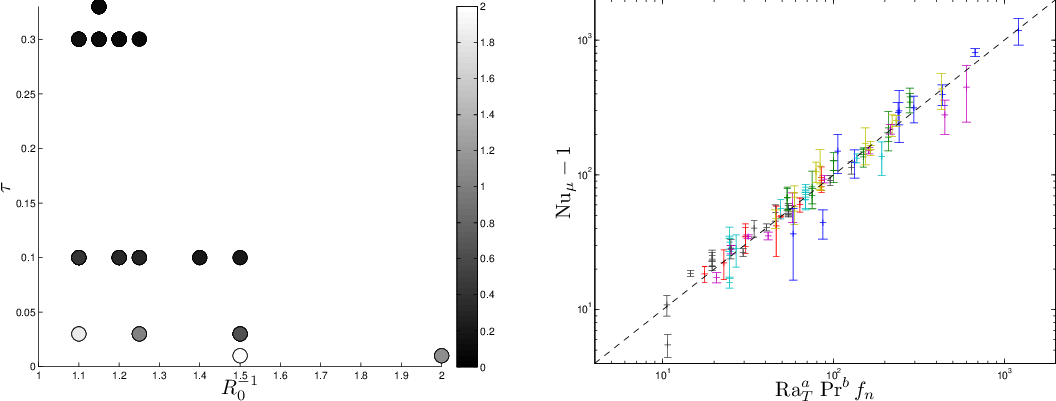}%
  \caption{\captionsize
    Left panel:
    The best-fit values for the coefficients $f_n$ defined in \eq(\ref{eq:NuC_guess2}).
    The color bar has been rescaled to accommodate the larger range of values here.
    Right panel:
    Comparison between the measured values of $\Nu_\mu - 1$
    and the best fit formula in \eq(\ref{eq:NuC_guess2}) with exponents $a=0.37$ and $b=0.27$.}
  \label{fig:NuC_fit}
\end{figure}

\Fig\ref{fig:NuC_fit} shows
the best fit values for the coefficients $f_n$ and a comparison between the measured values of $\Nu_\mu$
and the best fit for \eq(\ref{eq:NuC_guess2}).
As found previously for $\Nu_T$, there is a weak increase in $\Nu_\mu$ as $\Roi$ is made smaller,
\ie\ as the composition gradient is reduced.  Unlike $\Nu_T$, however, $\Nu_\mu$ depends strongly on
$\tau$, \ie\ on the compositional diffusivity.  In fact, we find that $\Nu_\mu$ is roughly inversely proportional to $\tau$,
indicating that the mean turbulent flux of composition $\left<\uz\mu\right>$
is approximately independent of compositional diffusivity in the parameter
regime considered here.

\subsubsection{The Buoyancy Flux Ratio ($\gamma^{-1}$)}
\begin{figure}[h!]
  \centering%
  \includegraphics[width=\textwidth]{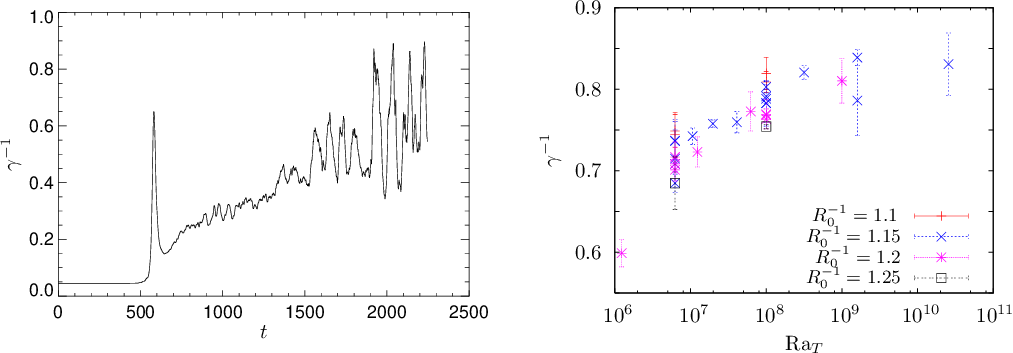}%
  \caption{\captionsize
    Left panel: Time series of $\gamma^{-1} = \tau\Roi\Mu_\mu/\Mu_T$
    from the same simulation shown in \fig\ref{fig:Nu}.  Right panel:
    The dependence of $\gamma^{-1}$ on $\Ra_T$ for all simulations with $\Pr=\tau=0.3$,
    and for the large-domain simulation shown in \fig\ref{fig:Nu_tall_box},
    which has $\Pr=\tau=0.33$.
    Different colors and symbols indicate different values of $\Roi$.}
  \label{fig:gamma}
\end{figure}

In all our simulations, we find that the flux ratio $\gamma^{-1}$ increases when layers
form, and increases still further when layers merge, so $\gamma^{-1}$
is an increasing function of $\Ra_T$.
For illustration, \fig\ref{fig:gamma} shows a time series
of $\gamma^{-1}$ from the same simulation presented in Section~\ref{sec:typical},
and also plots $\gamma^{-1}$ against $\Ra_T$ using data from several other simulations.
The fact that $\gamma^{-1}$ is an increasing function of $\Ra_T$
has important implications for the merging of layers, as we discuss in the next section.

Taken at face value,
the empirical power laws for $\Nu_T$ and $\Nu_\mu$ derived above
imply that $\gamma^{-1} \propto \Nu_\mu/\Nu_T \propto \Ra_T^{0.03}$.
However, as discussed in Section~\ref{sec:model},
$\gamma^{-1}$ cannot exceed unity in a statistically steady state,
and therefore cannot follow a simple power law in $\Ra_T$.
\Fig\ref{fig:gamma} suggests that $\gamma^{-1}$
asymptotes to a constant value at large values of $\Ra_T$.
Further simulations at high $\Ra_T$ are required to confirm this trend,
and to determine how this asymptotic value depends on the other parameters
$\Roi$, $\Pr$, and $\tau$.  \Fig\ref{fig:gamma} suggests that $\gamma^{-1}$ is a decreasing function of $\Roi$,
but again, more data is required to confirm this suggestion.

\section{SUMMARY AND DISCUSSION}
\label{sec:summary}
\subsection{Turbulent transport at $\Pr \ll 1$}
We have measured the
mean fluxes of heat and composition in
an extensive suite of numerical simulations of layered ODD convection.
In these simulations, layers develop spontaneously from an initially homogeneously turbulent state,
and subsequently merge into a single layer spanning the entire height of the simulation domain.
The formation and merging of layers is, to some extent,
a stochastic process, but nevertheless occurs at a similar rate in simulations with different
domain sizes and numerical resolutions.

The development of layers leads to a significant increase in the mean fluxes,
and the fluxes increase still further each time layers merge.
We find that the mean fluxes are approximately proportional to the average thickness of
the layers to the power $4/3$, as predicted by \citet{Turner65}.
The heat flux is independent of viscosity, but the composition flux is not.
Within the range of parameter values considered here, the thermal and compositional
Nusselt numbers are well approximated by
\begin{flalign}
  && \Nu_T - 1 &= A\,\Pr^{1/3}\,\Ra_T^{1/3}&
  \label{eq:empirical_NuT} \\
  &\mbox{and}& \Nu_\mu - 1 &= B\,\tau^{-1}\,\Pr^{1/4}\,\Ra_T^{0.37}&
  \label{eq:empirical_NuC}
\end{flalign}
where the prefactors $A$ and $B$ are weakly dependent on the parameters $\Roi$, $\tau$, $\Pr$, and $\Ra_T$,
and have typical values $A \simeq 0.1$ and $B \simeq 0.03$ respectively.
The mean fluxes of heat and composition can be calculated from these expressions using
\eqs(\ref{eq:flux_T}) and (\ref{eq:flux_C}).
We emphasize, however, that the parameter values used in our simulations
are rather far from true astrophysical values, and so any application of the formulae
(\ref{eq:empirical_NuT})--(\ref{eq:empirical_NuC})
to astrophysical conditions requires an extrapolation beyond the range in which these
formulae have been derived.
In particular, \fig\ref{fig:gamma}b suggests that the ratio of $\Nu_T$ and $\Nu_\mu$
tends to a constant for large values of $\Ra_T$, which is not compatible with the power law behavior assumed in equations (\ref{eq:empirical_NuT}) and (\ref{eq:empirical_NuC}).  Further work in larger computational domains
will be required to confirm this suggestion.

In all our simulations
the interfaces between the convective layers
exhibit complex dynamical behavior
that is very different from the laminar ``interface scouring'' and entrainment seen
in studies of layered convection at higher Prandtl number
\citep[\eg][]{LindenShirtcliffe78,Fernando89}.
The structure of the interfaces,
and their effective surface area,
is highly time dependent,
which produces order unity variations in the fluxes
of heat and composition
during each layered phase.
Previous models for the transport by ODD convection have assumed that the interfaces remain flat and laminar,
implying that the fluxes are limited by the vertical gradients of the horizontally averaged profiles
within the interfaces.
Our numerical results demonstrate that transport at low Prandtl number is more efficient than
predicted by such quasi-one-dimensional models.
We argue that the interface dynamics must be considered in future models for transport
under astrophysical parameter conditions.

\subsection{The Layer Thickness in Astrophysical Objects}
Since the transport by layered convection is strongly dependent on the layer thickness,
we must consider what determines this thickness in stellar and planetary interiors.
In all of our simulations,
the layers eventually merge into a single layer with the same height
as the computational domain,
so our results make no prediction for the thickness of layers in real objects.
The detailed dynamics of layer mergers will be the subject of a future publication,
but we now provide a qualitative comparison of our results with other models of layered convection.

The dynamics of layer mergers in
\emph{fingering} convection (that is, double-diffusive convection driven by a destabilizing
\emph{compositional} gradient in a subadiabatic region)
has been discussed by \citet{Radko05} in the context of the Earth's oceans.
Using a mean-field model for the turbulent transport of heat and composition,
\citeauthor{Radko05} argues
that layers stop merging once they achieve a saturation thickness, $H = H_0$, say.
Recently, \citet{LeconteChabrier12} have
applied \citeauthor{Radko05}'s arguments to ODD convection,
as part of a model for turbulent transport in Jupiter's interior.
They use heuristic arguments to obtain upper and lower bounds for $H_0$ in Jupiter,
and thereby derive bounds for the thermal and compositional transport.

For fingering convection, the roles of thermal and compositional stratification are reversed compared
to ODD convection.  Because \citeauthor{Radko05}'s model does not depend explicitly on the
microscopic diffusivities of temperature and composition,
his results can be translated directly to ODD convection
simply by replacing the density ratio $R_0$ and buoyancy flux ratio $\gamma$ by their respective
inverses $\Roi$ and $\gamma^{-1}$.
When thus applied to ODD convection,
his model makes two important assumptions about the turbulent transport in the layers and interfaces:
\begin{enumerate}
  \item Turbulent fluxes in the layers are described by an effective diffusivity
  that acts on the mean (horizontally-averaged) temperature and composition fields.
  The effective diffusivity
  is a function of the layer thickness $H$,
  and is the same for both temperature and composition.
  \item The flux ratio $\gamma^{-1}$ in the interfaces depends only on the \emph{local} mean value of the density ratio,
  $\RIi$ say, and the functional dependence of $\gamma^{-1}$ on $\RIi$
  is identical to
  the dependence of $\gamma^{-1}$ on $\Roi$ seen
  in homogeneous ODD convection.
  The qualitative form of this dependence was determined in \citetalias{Mirouh-etal12}:
  $\gamma^{-1}$ is a decreasing function of $\Roi$ for $1 < \Roi < \RLi$,
  and an increasing function of $\Roi$ for $\Roi > \RLi$, with a minimum at $\Roi = \RLi$.
\end{enumerate}
In a statistically steady state of layered ODD convection,
the fluxes through the layers and interfaces must be equal.
Using the first assumption, it can then be shown that
\begin{equation}
  \RIi - \Roi = (\Roi-\gamma^{-1})\frac{\Delta\vartheta_{\rm L}}{\Delta\vartheta_{\rm I}},
  \label{eq:jump}
\end{equation}
where $\Delta\vartheta_{\rm L}$ and $\Delta\vartheta_{\rm I}$ represent the jumps in
potential temperature across the layers and interfaces, respectively.
This result follows from \citeauthor{Radko05}'s equations (6b) and (7) and the definitions
of the density ratios in the layers and interfaces.
Since $\Roi > 1 >\gamma^{-1}$, it follows that the right-hand side of \eq(\ref{eq:jump}) is positive,
and therefore that $\RIi > \Roi$.
In other words, the interfaces must have a larger local value of $\Roi$ than the
homogeneous state of ODD convection from which they formed.
Since spontaneous layer formation occurs only if $\Roi < \RLi$, this generally implies,
using the second assumption,
that $\gamma^{-1}$ should decrease when layers form.
By making further assumptions about the turbulent fluxes,
\citet{Radko05} then showed that $\RIi$ continues to increase, and $\gamma^{-1}$
continues to decrease, each time layers merge, until $\RIi$ exceeds the critical value $\RLi$
corresponding to the minimum of $\gamma^{-1}$.
At this point the layer mergers cease,
and so the saturation layer thickness $H_0$
is determined by the condition $\RIi = \RLi$.

Although this picture of layer merging
is certainly plausible,
we find that it entirely conflicts with our numerical results.  In particular, we find in all of our simulations
that $\gamma^{-1}$ \emph{increases} when layers form, and increases still further when layers merge,
as shown in \fig\ref{fig:gamma}.
Therefore our simulations show that (at least) one of the two assumptions listed above must be incorrect.
In fact, we have already seen
that the fluxes through the interfaces
in our simulations
depend not only on the local temperature and composition gradients,
but also on the effective surface area of the interfaces, which we find to be highly time dependent.
This implies that \citeauthor{Radko05}'s second assumption is violated in our simulations,
and that a more reliable model for the merging of layers will need to account for the large-scale
deformations of the interfaces illustrated in \fig\ref{fig:min_max}.
We emphasize that \citeauthor{Radko05}'s model was derived in the context of oceanic fingering
convection, for which the interfaces do not exhibit the complex dynamical behavior seen here.\footnote{%
  Whether \citeauthor{Radko05}'s assumptions hold even for oceanic fingering convection is currently unknown.
  In that context, his model implies that $\gamma$ decreases when layers form, which seems to contradict
  the results of \citet[][their \fig3d]{Stellmach-etal11}.%
}

The incompatibility of our results with \citeauthor{Radko05}'s predictions
does not, however, rule out the existence of a finite saturation layer thickness.
In practice, the thickness of layers in real astrophysical objects may also be limited by factors
that are outside the scope of the numerical model presented here.
In particular,
the Boussinesq approximation
breaks down once the layer thickness becomes comparable to the local pressure scale height
of the planet or star.
At this point, convection within the layers would be almost indistinguishable from ``normal'' overturning
convection, and the convective transport could then be parameterized using the usual mixing-length approximation.
Our model also assumes a prescribed difference in temperature and composition between the top and bottom
of the numerical domain, rather than a prescribed flux, and so there is no imposed upper limit on the turbulent
fluxes in our simulations.  In a real physical system, on the other hand, these fluxes
are limited by the global heat and composition budgets,
and by the modification of the background state by the turbulent fluxes.
It is conceivable that the merging of layers
will cease once the convection becomes efficient enough to transport all of the available heat and composition,
or once the back reaction of the fluxes on the overall gradients becomes significant.
These and other constraints on the layer thickness have been discussed by \citet{LeconteChabrier12}.

Finally, an important issue that has not been explicitly addressed so far is the timescale for layer merging.
Our results suggest that this timescale is an increasing function of layer height,
which is in accordance with the theory of layer mergers proposed by \citet{Radko05}.
If the timescale for merging in an astrophysical object eventually becomes comparable to the
global evolution timescale, then the layers will never reach a saturation thickness,
and layered convection will persist throughout the object's evolution.
In that case,
the evolution of the layers must be modeled as an explicitly time-dependent
process in 1D structure and evolution models.
Future papers will study the dynamics of layer mergers in more detail,
in order to determine what sets the merging timescale, and how it depends on layer height.

We thank G.~Chabrier, B.~R.~Ruddick, E.~A.~Spiegel, and M.~G.~Wells
for useful comments and suggestions.
P.~G. and T.~W. were supported by funding from the NSF (NSF-0933759).
Part of the computations were performed on the UCSC Pleiades supercomputer,
purchased with an NSF-MRI grant.
Others used computer resources at the National Energy Research Scientific Computing Center (NERSC),
which is supported by the Office of Science of the US Department of Energy
under contract DE-AC03-76SF00098.

\appendix
\section{INFLUENCE OF DOMAIN SIZE AND NUMERICAL RESOLUTION}
\label{sec:robust}

In this appendix we investigate the influence of
numerical resolution and domain size on
the initial layer height,
the timescale for layer formation and merging,
the time-averaged fluxes,
and
the amplitude of oscillations in the fluxes.

\subsection{Tests of Numerical Resolution}

Because of the disparity between the diffusivities of temperature and composition,
double-diffusive convection always features a
wide range of dynamical length scales,
making this a tough problem to model numerically.
Furthermore,
as mentioned in Section~\ref{sec:typical},
the emergence of layers is associated with an increase in the overall level of turbulence,
and increased spectral energy at all scales in a given simulation
\citep{Stellmach-etal11}.
Properly resolving the layered phase therefore
requires even greater numerical resolution than resolving the homogenous phase.
In order to check that the thermal and compositional fluxes in our simulations
are not sensitive to the numerical resolution, we have run several
simulations that have the same physical parameters and domain size, but different
numbers of grid points
in the vertical and horizontal directions.

\begin{figure}[h!]
  \centering%
  \includegraphics[width=12cm]{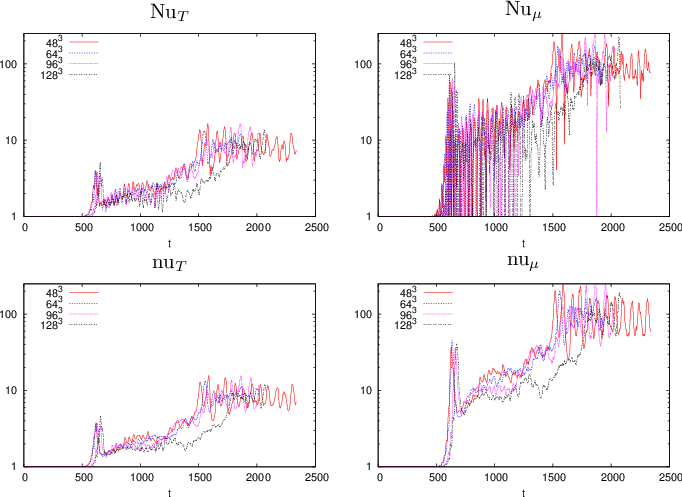}%
  \caption{\captionsize
    Comparison of Nusselt numbers between simulations
    with $48^3$, $64^3$, $96^3$, and $128^3$ Fourier modes in each spatial direction.
    All four cases have $\Pr = \tau = 0.03$, $R_0^{-1} = 1.5$,
    and a domain size of $(50d)^3$.
    The two rows show the Nusselt numbers
    calculated from the instantaneous fluxes (\ref{eq:NuT}) and (\ref{eq:NuC}),
    and the instantaneous dissipation rates (\ref{eq:MuT}) and (\ref{eq:MuC}).}
  \label{fig:res_check}
\end{figure}

As an example,
\fig\ref{fig:res_check} compares
time series of $\Nu_{T,\mu}$ and $\Mu_{T,\mu}$
from four simulations performed
with different numerical resolutions.  All four were performed in domains
of size $(50d)^3$ with the same physical parameters as the simulation presented in Section~\ref{sec:typical},
but with $48^3$, $64^3$, $96^3$, and $128^3$ Fourier modes respectively.
The simulation with the coarsest resolution ($48^3$)
only resolves structures on lengthscales $\gtrsim d$, but even in this simulation
the linear double-diffusive instability is well resolved (recall that the fastest growing linear mode has
wavelength $\simeq 20d$ at these parameters).
Each simulation undergoes a homogeneous phase, then forms three layers,
which later merge into two layers, then one layer.
The duration of each layered phase varies between the four simulations,
suggesting that the merging of layers is, to some extent, a stochastic process.
However, the four simulations produce similar fluxes during each phase,
and the variations in each flux about its mean are comparable in all four cases.

\begin{figure}
  \centering%
  \includegraphics[width=16cm]{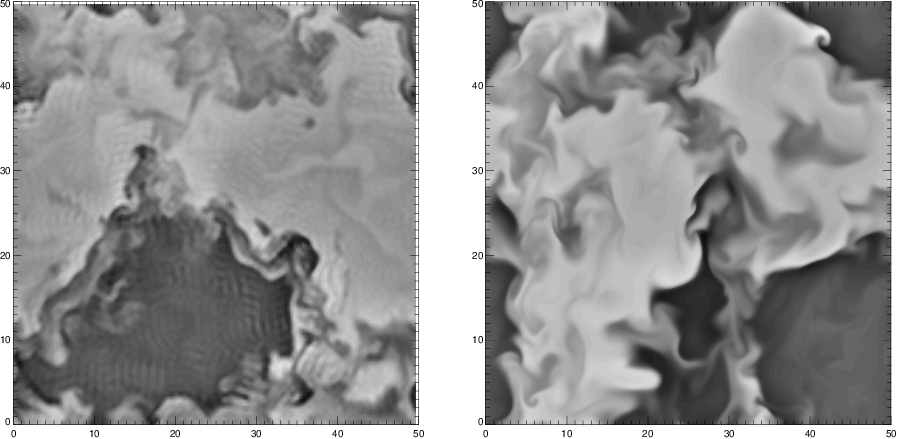}%
  \caption{Horizontal cross-sections of density $\rho$ from the $48^3$ resolution simulation at $t=1530$ (left)
  and the $128^3$ resolution simulation at $t=1800$ (right).
  Both cases show cross-sections through the interface at the start of the one-layer phase,
  which is typically the least well-resolved stage in any simulation.}
  \label{fig:high_low}
\end{figure}

In each simulation, the most stringent test of the numerical resolution occurs during the final merging event,
which marks the start of the one-layer phase.
\Fig\ref{fig:high_low} compares the density fields at this point
in the lowest and highest resolution simulations shown in \fig\ref{fig:res_check}.
The lowest resolution simulation is clearly under-resolved, with significant features at the Nyquist scale,
whereas the highest resolution simulation displays no numerical artifacts.
Nevertheless, the mean fluxes through these two simulations
are approximately equal in the one-layer phase, which suggests
that the transport of both heat and composition is dominated
by scales $\gtrsim d$ in these simulations.

\subsection{Tests of Domain Size and Aspect Ratio}

\begin{figure}[h!]
  \centering%
  \includegraphics[width=16cm]{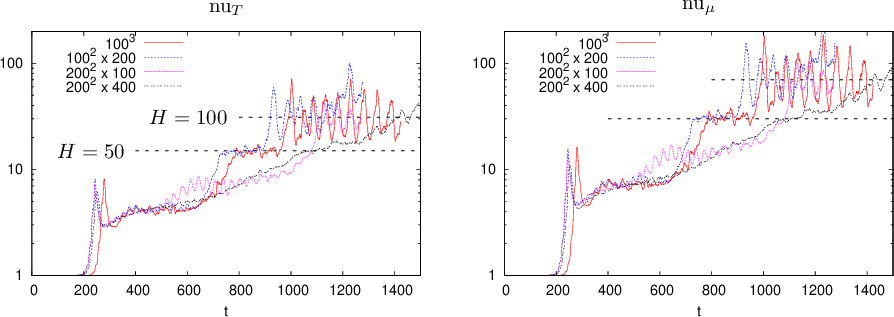}%
  \caption{\captionsize
    Comparison of Nusselt numbers between simulations in domains of different sizes.
    All four cases have $\Pr = \tau = 0.3$ and $\Roi=1.15$.
    The dashed lines indicate the mean values of $\Mu_T$ and $\Mu_\mu$
    for layer heights of $H=50d$ and $H=100d$.}
  \label{fig:4cases}
\end{figure}

\Fig\ref{fig:4cases} compares time series of the Nusselt numbers from four simulations
performed in domains of different heights and widths.  Each simulation
has $\Pr=\tau=0.3$, $\Roi=1.15$, and the same numerical resolution,
\ie\ the same (dimensionless) distance between neighboring grid points.
These four simulations exhibit the following features, which we find to be typical:
\begin{itemize}
  \item The cases with wider domains ($200d$) take longer to form layers
    than the cases with narrower domains ($100d$).
  \item The initial layer height is not very sensitive to the size of the domain,
    provided that the domain is sufficiently wide.
    In \fig\ref{fig:4cases}, for example, the initial layer height is $50d$ in all four cases.
    However, in other simulations
    (not shown here, but see \citet{Rosenblum-etal11} for instance)
    it has been found that in very narrow domains
    layers form much earlier, and the initial layer height is smaller.  This result indicates
    that the horizontally-periodic boundary conditions, which encourage horizontal coherence
    in the simulation,
    can artificially accelerate the formation of layers when the domain is sufficiently narrow.
    For this reason, we only use data from simulations that were performed in domains of width $\geqslant100d$
    when calculating the mean fluxes in Section~\ref{sec:measure}.
  \item The mean fluxes of heat and composition, for layers of a particular height,
    are consistent between runs with different domain sizes.
  \item During each layered phase, the cases with wider domains exhibit smaller variations in the value of
  each flux about its mean value.  This suggests that the variations in the fluxes are horizontally localized,
  and therefore have less effect on the mean flux in wider domains.
  Similarly, the simulation in the tallest domain (of height $400d$) exhibits the smallest variations in the fluxes
  during each layered phase, because in this simulation the volumetric mean flux represents
  an average over a greater number of layers and interfaces.
\end{itemize}

\Fig\ref{fig:4cases} also illustrates a practical difficulty that arises when comparing results
from simulations performed in domains of different heights,
which is that simulations in taller domains usually undergo more layered phases
than simulations in shorter domains, and the length of each layered phase is correspondingly shorter.
For example,
the simulation with a domain height of $400d$
passes through phases in which the average layer height is $400d/7$ and $400d/6$;
these states cannot be achieved in the simulations with a domain height of $100d$.
As a result, the time series of the Nusselt numbers in the taller simulation are less obviously
step-like, and the mean flux during each layered phase is more difficult to measure accurately.
This also makes it difficult to determine whether the domain height affects the rate at which layers merge.
The full time series of both Nusselt numbers from this simulation are shown in \fig\ref{fig:Nu_tall_box}a.
We find that the time between layer merger events increases as the number of layers decreases,
which is in accordance with the model of layer merging proposed by \citet{Radko05}.

Another difficulty associated with measuring fluxes in simulations with multiple layers is that the layers
are not necessarily all of the same height.  In that case, a complete description of the state
of the system would require (at least) measuring the height of each layer, and the jumps in
temperature and composition across each interface, potentially introducing many new parameters
on which the mean flux might depend.
Furthermore, in such cases we would not expect the system to be in statistical equilibrium,
and so the mean flux might not be well defined.
To avoid dealing with this additional complexity, in this study
we assume that, in each layered phase, all the layers have approximately the same height,
while recognizing that this assumption introduces an additional source of error into our flux measurements.

\end{document}